\documentstyle[aps,preprint,12pt]{revtex}
\begin{document}
\preprint{hep-th/9611236}
\title{A Unified Treatment of the Characters of SU(2) and SU(1,1)}
\author{Subrata Bal\footnote{Presently at The Institute of Mathematical Sciences,  CIT Campus, 
Madras - 600 113,  India, Email-subrata@imsc.ernet.in .} 
, K. V. Shajesh \footnote{Presently at Physical Research Laboratory, Navarangapura, Ahamedabad - 380 009, India,\\         Email-kvs@prl.ernet.in .}  and Debabrata Basu.}
\address{Department of Physics \\Indian Institute of Technology \\ Kharagpur- 721 302, West Bengal,
 India.}
\maketitle

\baselineskip 27pt 

\begin{abstract}

     The character problems of SU(2) and SU(1,1) are reexamined from the
standpoint of a physicist by employing the Hilbert space method which is
shown to yield a completely unified treatment for SU(2) and the discrete
series of representations of SU(1,1). For both the groups the problem is 
reduced to the
evaluation of an integral which is invariant under rotation for SU(2) and
Lorentz transformation for SU(1,1). The integrals are accordingly
evaluated by applying a rotation to a unit position vector in SU(2) and a
Lorentz transformation to a unit SO(2,1) vector which is time-like for the
elliptic elements and space-like for the hyperbolic elements in SU(1,1).
The details of the procedure for the principal series of representations of
SU(1,1) differ substantially from those of the discrete series.

\end{abstract}

\section{Introduction}\label{i}

     A major tool in group representation theory is the theory of character.
The importance of the concept of character of a representation stems
from the fact that for a semisimple Lie group every unitary irreducible
representation is uniquely determined by its character.  The
simplification effected by such an emphasis is obvious; in particular the
formal processes of direct sum and direct product as applied to
representations are reflected in ordinary sum and multiplication of
characters.

     For finite dimensional representations character is traditionally
defined as the sum of the eigenvalues of the representation matrix .
It should be pointed out that the unitary operators of an infinite
dimensional Hilbert space do not have character in this sense since the
infinite sum consists of numbers of unit modulus.  For example for $g=e$ one
has $D(e)=I$ and the sum of the diagonal elements of the infinite
dimensional unit matrix is $\infty$.  We now briefly give the
Gel'fand-Naimark $\!\!^1$ definition of character which introduces the concept
through the group ring as a generalized function on the group manifold.

     We denote by $X$ the set of infinitely differentiable functions
$x(g)$ on the group, which are equal to zero outside a bounded set.  If 
$g\rightarrow T_{g}$ be a representation of
the group $G$ we set 
\begin{equation}\label{iGRNGd} T_{x}=\int
     x(g)T_{g}d\mu(g) 
\end{equation} 
where $d\mu(g)$ is the left and right invariant measure (assumed
coincident) on $G$ and the integration extends over the entire group manifold.

     The product $T_{x_{1}}T_{x_{2}}$ can be written in the form,
\[T_{x_{1}}T_{x_{2}}=\int x(g)T_{g}d\mu(g)\]
where
\begin{equation}\label{ifnc}
x(g)=\int x_{1}(g_{1})x_{2}(g_{1}^{-1}g)d\mu(g_{1})
\end{equation}
The function $x(g)$ defined by equation (~\ref{ifnc}) will be called the
product of the functions  $x_{1}$, $x_{2}$ and denoted by $x_{1}x_{2}(g)$.

     Let us suppose that $g\rightarrow T_{g}$ is a unitary representation
of the group $G$ realized in the Hilbert space $H$ of the functions $f(z)$
with the scalar product
\[(f,g)=\int \overline{f(z)}g(z) d\lambda(z)\]
where $d\lambda(z)$ is the measure in $H$.

     Then the operator $T_{x}$ is an integral operator with a kernel:
\[
T_{x}f(z)=\int K(z,z_{1})f(z_{1})d\lambda(z_{1})
\]
It then follows that $K(z,z_{1})$ is a positive definite Hilbert-Schmidt
kernel, satisfying
\[ \int |K(z,z_{1})|^{2}d\lambda(z)d\lambda(z_{1})<\infty\]
Such a kernel has a trace $Tr(T_{x})$ where
\[
Tr(T_{x})=\int K(z,z)d\lambda(z)
\]
Using the definition in equation (~\ref{iGRNGd}) one can prove that
$Tr(T_{x})$ can be written in the form
\begin{equation}\label{iDtrtx} Tr(T_{x})=\int
     x(g)\pi(g)d\mu(g) 
\end{equation} 

     The function $\pi(g)$ is the character of the 
representation $g \rightarrow T_{g}$.  It should be noted that in this
definition the matrix representation of the group does not appear and, as
will be shown below, it makes a complete synthesis of the finite and
infinite dimensional irreducible unitary representations.

     The character of the complex and real unimodular groups was evaluated
by Gel'fand and coworkers $\!\!^{2,3}$.  The real group SL(2,R) turns out to be more
involved than the complex group particularly because
of the presence of the
discrete series of unitary irreducible representations (unirreps).  The main problem
in the Gel'fand-Naimark theory of character is the construction of the
integral kernel $K(z,z_{1})$ which requires a judicious choice of the
carrier space of the representation.  The representations of the positive
discrete series $D_{k}^{+}$ were realised by Gel'fand and
coworkers $\!\!^{3}$ in the space of the functions on the half-line $R^{+}$ and
those of the negative discrete series $D_{k}^{-}$ on the half-line
$R^{-}$.  The integral kernel of the group ring was determined by them
essentially for the reducible representation $D_{k}^{+}\oplus D_{k}^{-}$ which
considerably complicates the subsequent computation of the character of a
single irreducible representation.  It is the object of this paper to
re-examine the character problem of SU(1,1) (or SL(2,R)) from a
physicist's standpoint by using the Hilbert space method developed by
Bargmann $\!\!^{4}$ and Segal $\!^{5}$ in which computations can be carried out within a
single unirrep  of the positive or negative discrete
series. This method not only simplifies the crucial problem of
construction of the integral kernel of the group ring but serves as the
key to the synthesis of the finite and infinite dimensional
representations mentioned above. The Hilbert space method as applied to
the evaluation of the characters of SU(2) and SU(1,1) proceeds along
entirely parallel lines. For SU(2) the problem essentially reduces to the
evaluation of an integral of the form 
\begin{equation}\label{iCs}
\int_{0}^{\pi} d\theta \sin\theta \int_{0}^{2\pi}d\phi\left( \cos\frac{\theta_{0}}{2}
-i\sin\frac{\theta_{0}}{2}\hat{n}.\hat{r}\right)^{2j}
\end{equation}
where $\hat{n}$ and $\hat{r}$ are unit position
vectors, $\hat{n}$ being fixed and $\hat{r}(\theta,\phi)$ being
the variable of integration. This integral is easily evaluated
by rotating the co-ordinate axes such that the 3-axis (Z-axis)
coincides with $\hat{n}$

     The character of the elliptic elements of SU(1,1) is given by an
integral closely resembling the above:
\begin{equation}\label{iCup}
     \int_{\tau=0}^{\infty} d\tau\; \sinh\tau 
 \int_{0}^{2\pi}d\phi \left[ \cos\frac{\theta_{0}}{2}
 -i\sin\frac{\theta_{0}}{2} \hat{n}. \hat{r} \right]^{-2k}  
\end{equation}
where $\hat{n}$ and $\hat{r}$ are a pair of unit time-like SO(2,1)
vectors, $\hat{n}$, as before, being fixed and $\hat{r}$ the variable of
integration. This integral is evaluated by an appropriate Lorentz
transformation such that the time axis points along the fixed time-like
SO(2,1) vector $\hat{n}$. For the hyperbolic elements of SU(1,1) the above
integral is replaced by
\[ \int_{0}^{\infty}
     d \tau \sinh \tau \int d\phi\left[\epsilon\cosh\frac{\sigma}{2}-
     i\sinh\frac{\sigma}{2}\hat{n}.\hat{r}\right]^{-2k} \]
     where $\epsilon=\underline{+}1$, $\hat{n}$ is a unit space-like and
$\hat{r}$ is a unit time-like SO(2,1) vector.  This integral is once again
evaluated by a Lorentz transformation such that the first space axis
(X-axis) coincides with the fixed space-like vector $\hat{n}$.  The
explicit evaluation is, however, a little lengthier than that for the elliptic
elements.

For the principal series of representations the carrier space is chosen to be 
the traditional Hilbert space $\!\!^{6}$ of functions defined on the unit 
circle.  Although the broad outlines of the procedure is the same as above the 
details differ substantially from those of the discrete series. An important 
feature of the principal series of representations is that the elliptic elements
of SU(1,1) do not contribute to its character.

\section{The Group SU(2) and The Discrete Series of Representations of SU(1,1)}

To make this paper self contained we describe the basic properties of the Hilbert 
spaces of analytic functions for SU(2) and SU(1,1).

\subsection{The Group SU(2) }

 The group SU(2) consists of 2X2 unitary, unimodular matrices 
\begin{equation}\label{sdSD}
     u=\left(\begin{array}{cc}
        \alpha &\beta\\
        -\bar{\beta} &\bar{\alpha}
     \end{array}\right) , \hspace{.5 in} |\alpha|^{2}+|\beta|^{2}=1
\end{equation}

 We know that every unitary, unimodular matrix $u$ can be
 diagonalized by a unitary unimodular matrix $v$ so that 
\begin{equation}\label{sdPmeter}
     u= v \epsilon(\theta_{0}) v^{-1}
\end{equation}
where $\epsilon(\theta_{0})$ is the diagonal form of $u$ :
\[
     \epsilon(\theta_{0}) =\left(\begin{array}{cc}
     e^{\frac{i\theta_{0}}{2}} &0\\
     0 &e^{-\frac{i\theta_{0}}{2}}\end{array}\right)\]
  Since $v$ is also an SU(2) matrix it can be factorized in terms of Euler
 angles as, 
\begin{equation}\label{spfac}
     v=\epsilon(\eta)a(\tau)\epsilon(\chi),
\end{equation}
where 
\[
     a(\tau)=\left(\begin{array}{cc}
     \cos\frac{\tau}{2} &\sin\frac{\tau}{2}\\
      -\sin\frac{\tau}{2} &\cos\frac{\tau}{2}
     \end{array}\right)\]
We therefore obtain the following parametrization of $u$ : 
\begin{equation}\label{Spara}
u = \!\epsilon(\eta)a(\tau)\epsilon(\theta_{0})a^{-1}(\tau)\epsilon^{-1}(\eta).
\end{equation}
The parametrization (~\ref{Spara}) yields
\begin{eqnarray}\label{sdPmtr}
     \alpha &= &\cos\frac{\theta_{0}}{2}+i\sin\frac{\theta_{0}}{2}\cos\tau
     \\
     \beta &= &-ie^{i\eta}\sin\frac{\theta_{0}}{2}\sin\tau
\label{sdPmtr1}
\end{eqnarray}

   The representations of SU(2) will be realized in the Bargmann-Segal 
 space $B(C_{2})$ which consists of entire analytic functions $\phi(z_{1},z_{2})$
 where $z_{1}$ and $z_{2}$ are spinors transforming according to the fundamental
 representation of SU(2) :  
\begin{equation}\label{srGRactn}
     (z_{1}^{\prime},z_{2}^{\prime})=(z_{1},z_{2})u
\end{equation}     
 The action of the finite element of the group in $B(C_{2})$ is given by,
 \[
     T_{u}\phi(z_{1},z_{2})=\phi(\alpha z_{1} - \bar{\beta} z_{2}, 
     \beta z_{1} + \bar{\alpha} z_{2})
\]
To decompose $B(C_{2})$ into the direct sum of the subspaces $B_{j}(C)$ 
invariant under SU(2) we introduce Schwinger's angular momentum operators 
in $B(C_{2})$
 \begin{eqnarray*}
      J_{1} &= &\frac{1}{2}\left(z_{1}\frac{\partial}{\partial
     z_{2}}+z_{2}\frac{\partial}{\partial z_{1}}\right)
     \\ 
      J_{2} &= &-\frac{i}{2}\left(z_{1}\frac{\partial}{\partial
     z_{2}}-z_{2}\frac{\partial}{\partial z_{1}}\right)
     \\ 
      J_{3} &= &\frac{1}{2}\left(z_{1}\frac{\partial}{\partial
     z_{1}}-z_{2}\frac{\partial}{\partial z_{2}}\right) 
\end{eqnarray*}
Explicit calculation yields

\[ {\vec{J}}^{2}=J_{1}^{2}+J_{2}^{2}+J_{3}^{2}=\frac{K}{2}\left(\frac{K}{2}+1 \right) \]     
where $K$ stands for the operator
\[ K=\left(z_{1}\frac{\partial}{\partial
     z_{1}}+z_{2}\frac{\partial}{\partial z_{2}}\right)\]     
 Since $K$ commutes with all the components of the angular momentum in  
 an irreducible representation it can be replaced by 2j, j = 0,
 $\frac{1}{2}$, 1, $\frac{3}{2}$, 2,.....The subspace $B_{j}(C)$ is, therefore,
 the space of homogeneous polynomials of degree 2j in $z_{1}$, $z_{2}$ :
\begin{equation}\label{spol}
\phi(z_{1},z_{2})=[(2j)!]^{-\frac{1}{2}}z_{2}^{2j}f(z), \vspace{.5 in}
z=\frac{z_{1}}{z_{2}}
\end{equation} 
where the numerical factor $[(2j)!]^{-\frac{1}{2}}$ is introduced for 
convenience. If we restrict ourselves to functions of the form (~\ref{spol}) the finite 
element of the group is given by 
\begin{equation}\label{srGRre}
     T_{u}f(z)=(\beta z + \bar{\alpha})^{2j} f \left(
     \frac{\alpha z - \bar{\beta}}{\beta z + \bar{\alpha}} \right)
\end{equation}
This representation is unitary with respect to the scalar product 
 \begin{equation}\label{srpDEfn}
     (f,g)=\int \overline{f(z)}g(z) d\lambda(z)
\end{equation} 
where
\begin{equation}\label{srpLAM}
     d\lambda(z)=\frac{(2j+1)}{\pi}(1+|z|^{2})^{-2j-2}d^{2}z,
\end{equation}
\[d^{2}z= dx dy, \vspace{.5 in} z= x + iy\]
The principal vector in this space is given by 
 \begin{equation}\label{srvPvec}
     e_{z}(z_{1})=(1+\bar{z}z_{1})^{2j}
\end{equation}     
 so that 
\begin{equation}\label{prnv} 
     f(z)= \int (1+z\bar{z}_{1})^{2j}f(z_{1})d\lambda(z_{1})
\end{equation}     
 We now construct the group ring which consists of the operators
\[ T_{x}=\int x(u) T_{u} d\mu(u) , \]
 where $d\mu(u)$ is the invariant measure on SU(2) and $x(u)$ is an 
 arbitrary test function on the group, which vanishes outside a bounded set.
 The action of the group ring is  then given by
 \[
     T_{x}f(z)=\int_{}^{} x(u)(\beta z+\bar{\alpha})^{2j}f \left( \frac{\alpha
     z-\bar{\beta}}{\beta z+\bar{\alpha}} \right)d\mu(u) 
\]
  We now use the reproducing kernel as given by (~\ref{prnv}) to write
  \[
     f \left( \frac{\alpha  z-\bar{\beta}}{\beta z+\bar{\alpha}} \right)=
     \int \left[ 1+ \frac{(\alpha z-\bar{\beta})\bar{z}_{1}}{(\beta
     z+\bar{\alpha})} \right]^{2j} f(z_{1})d\lambda(z_{1})\]
  Thus
 \[
     T_{x}f(z)=\int K(z,z_{1})f(z_{1})d\lambda(z_{1}) \]     
where the kernel $K(z,z_{1})$ is given by,
\begin{equation}\label{scK}
     K(z,z_{1})= \int x(u) (\beta z+ \bar{\alpha})^{2j} \left[ 1+ \frac{(\alpha z-\bar{\beta})\bar{z}_{1}}{(\beta
     z+\bar{\alpha})} \right]^{2j} d\mu(u)
\end{equation}     
 Since the kernel $K(z,z_{1})$ is of the Hilbert-Schmidt type we have
\[ Tr(T_{x})= \int K(z,z)d\lambda(z) \]     
Using the definition (~\ref{scK}) of the kernel we have 
\[ Tr(T_{x})= \int x(u)\pi(u)d\mu(u) \]
where
 \begin{equation}
     \pi(u) = \int (\beta z+ \bar{\alpha})^{2j} \left[ 1+ 
 \frac{(\alpha z-\bar{\beta})\bar{z}}{(\beta
     z+\bar{\alpha})} \right]^{2j} d\lambda(z)
\label{scC1}
\end{equation}
Setting 
 \[ z= \tan\frac{\theta}{2}e^{i\phi}, \hspace{ .5 in} 
 0 \leq \theta <\pi ,0 \leq \phi \leq 2\pi\]
and using the parametrization (~\ref{sdPmtr},\ref{sdPmtr1}) we obtain after some calculations,
\begin{equation}\label{scC5}
     \pi(u) =\!\frac{2j+1}{4\pi} \int_{\theta=0}^{\pi} \! d\theta \sin\theta 
 \int_{0}^{2\pi}\!d\phi \left[ \cos\frac{\theta_{0}}{2}
 -i\sin\frac{\theta_{0}}{2} (\cos\tau\cos\theta +
 \sin\tau\sin\theta\cos\phi) \right]^{2j}       
\end{equation}
If we now introduce the unit vectors $\hat{n}$ and $\hat{r}$ as
 \[\hat{n}=(\sin\tau,0,\cos\tau), \hat{r}=(\sin\theta\cos\phi
 ,\sin\theta\sin\phi,\cos\theta)\]
the equation (~\ref{scC5})    can be written as
\begin{equation}\label{scC6}
     \pi(u) =\frac{2j+1}{4\pi} \int_{\theta=0}^{\pi} d\theta \sin\theta 
 \int_{0}^{2\pi}d\phi \left[ \cos\frac{\theta_{0}}{2}
 -i\sin\frac{\theta_{0}}{2} \hat{n}. \hat{r} \right]^{2j}   
\end{equation}
We now rotate the coordinate system such that the 3-axis (Z-axis) coincides with
the fixed vector $\hat{n}$. Thus
\[\pi(u)=\frac{2j+1}{2} \int_{\theta=0}^{\pi}  
  \left[ \cos\frac{\theta_{0}}{2}
 -i\sin\frac{\theta_{0}}{2} \cos\theta \right]^{2j} \sin\theta d\theta\]   
 The above integral is quite elementary and yields
 \[\pi(u)=\frac{\sin(j+\frac{1}{2})\theta_{0}}{\sin\frac{\theta_{0}}{2}}\]

\subsection{The Group SU(1,1)}     
The group SU(1,1) consists of pseudo-unitary, unimodular matrices 
\begin{equation}\label{udUD}
     u=\left(\begin{array}{cc}
        \alpha &\beta\\
        \bar{\beta} &\bar{\alpha}
     \end{array}\right),\hspace{.5 in}
\det(u) =|\alpha|^{2}-|\beta|^{2}=1
\end{equation}
and is isomorphic to the group SL(2,R) of real unimodular matrices,
\begin{equation}\label{udHslD}
     g= \left(\begin{array}{cc} a &b\\c &d \end{array} \right);\;
     \det(g)=ad-bc=1
\end{equation}
 A particular choice of the isomorphism kernel is 
\begin{equation}\label{udISOslm}
     \eta=\frac{1}{\sqrt{2}}\left( \begin{array}{cc}
     1 &i\\ i &1 \end{array}\right)
\end{equation}
so that 
\[
  u=\eta g \eta^{-1}
\]
\begin{equation}\label{udUSL}
     \alpha=\frac{1}{2}[(a+d)-i(b-c)];\;
     \beta= \frac{1}{2}[(b+c)-i(a-d)]
\end{equation}
The elements of the group SU(1,1) may be divided into three subsets : a)
elliptic, b) hyperbolic and c) parabolic. We define them as follows. 
Let $\alpha=\alpha_{1}+i\alpha_{2}$ and $\beta=\beta_{1}+i\beta_{2}$
so that
\[\alpha_{1}^{2}+\alpha_{2}^{2}-\beta_{1}^{2}-\beta_{2}^{2}=1\]
The elliptic elements are those for which 
\[\alpha_{2}^{2}-\beta_{1}^{2}-\beta_{2}^{2}>0\]
Hence if we set \[
\alpha'_{2}=\sqrt{\alpha_{2}^{2}-\beta_{1}^{2}-\beta_{2}^{2}} \]
we have
\[ \alpha_{1}^{2}+\alpha_{2}^{\prime^{2}}=1 \]
 so that $-1<\alpha_{1}<1$.  

     On the other hand the hyperbolic elements of SU(1,1) are those for 
which
\[\alpha_{2}^{2}-\beta_{1}^{2}-\beta_{2}^{2}<0\]
Hence if we write
\[ \alpha'_{2}=\sqrt{\beta_{1}^{2}+\beta_{2}^{2}-\alpha_{2}^{2}} \]
we have 
\begin{equation}\label{uhyp}
\alpha_{1}^{2}-\alpha_{2}^{\prime^{2}}=1
\end{equation}
 so that $|\alpha_{1}|>1$.

     We exclude the parabolic class corresponding to
\[\alpha_{2}=\sqrt{\beta_{1}^{2}+\beta_{2}^{2}}\]
as this is a submanifold of lower dimensions.

     If we diagonalize the SU(1,1) matrix (~\ref{udUD}), the eigenvalues
are given by 
\[\lambda=\alpha_{1}\underline{+}\sqrt{\alpha_{1}^{2}-1}\]
We shall consider the elliptic case $-1 < \alpha_{1} < 1 $ first. Thus,
setting $\alpha_{1} = \cos(\frac{\theta_{0}}{2}), 0<\theta_{0}<2\pi$
we have $\lambda = exp( \underline{+} i\frac{\theta_{0}}{2})$. We shall now show
 that every elliptic element of SU(1,1) can be diagonalized by a
pseudounitary transformation $ie$ 
\begin{equation}\label{udPmeter}
     v^{-1}uv= \epsilon(\theta_{0}), \hspace{.4 in}
     \epsilon(\theta_{0})=\left(\begin{array}{cc}
     \delta_{1} &0\\
     0 &\delta_{2}\end{array}\right),
\hspace{.4 in}
\delta_{1}=\bar{\delta_{2}}=e^{i\frac{\theta_{0}}{2}},
\end{equation}
 where  $v\in$ SU(1,1)\\ 
To prove this we first note that equation (~\ref{udPmeter}) can be written
as 
\begin{equation}\label{udEvec}
     uv_{1}=\delta_{1}v_{1}, uv_{2}=\delta_{2}v_{2}
\end{equation}
 where 
\[   v_{1}=\left(\begin{array}{c} v_{11}\\v_{21}\end{array}\right),
     v_{2}=\left(\begin{array}{c} v_{12}\\v_{22}\end{array}\right) \] 
Thus $v_{1}$ and $v_{2}$ are the eigenvectors of the matrix $u$
 belonging to the eigenvalues $\delta_{1}$ and $\delta_{2}$ respectively.  Hence
 $v_{1}$ and $v_{2}$ are linearly independent so that $\det(v) \neq 0$.
 We normalize the matrix $v$ such that
 \[\det(v)=v_{11}v_{22}-v_{12}v_{21}=1\]
 We now show that the eigenvectors $v_{1}$ and $v_{2}$ are 
 pseudoorthogonal $ie$ orthogonal with respect to the metric
     \[\sigma_{3}=\left(\begin{array}{cc} 1 &0\\ 0 &-1
     \end{array}\right)\]
In fact from the equation (~\ref{udEvec})  we easily obtain
\[\delta_{1}^{2}v_{2}^{\dagger}\sigma_{3}v_{1} =
     v_{2}^{\dagger}u^{\dagger}\sigma_{3}uv_{1}\]
Using the pseudounitarity of the matrix $u \in SU(1,1)$ we immediately
obtain 
 \[(v_{2}^{\dagger}\sigma_{3}v_{1})=0\]
If we further normalize 
\[v_{1}^{\dagger}\sigma_{3}v_{1}=1,\]
we easily deduce
 \[ v_{21}=\bar{v}_{12}, v_{22}=\bar{v}_{11}\]
Thus for the elliptic elements of SU(1,1) the transformation matrix $v$
defined by (~\ref{udPmeter}) is also an SU(1,1) matrix. Since every matrix $v \in$
SU(1,1) can be written as
\[v=\epsilon(\eta)a(\sigma)\epsilon(\theta),\]
where 
\begin{equation}\label{ast}
a(\sigma)=\left(\begin{array}{cc}
     \cosh\frac{\sigma}{2} &\sinh\frac{\sigma}{2}\\
      \sinh\frac{\sigma}{2} &\cosh\frac{\sigma}{2}
     \end{array}\right)\end{equation}
we immediately obtain 
\[
u
=\!\epsilon(\eta)a(\sigma)\epsilon(\theta_{0})a^{-1}(\sigma)\epsilon^{-1}(\eta)
\]
The above parametrization yields
\begin{eqnarray}\label{udPmtre}
     \alpha &= &\cos\frac{\theta_{0}}{2}+i\sin\frac{\theta_{0}}{2}\cosh\sigma
     \\
     \beta &= &-ie^{i\eta}\sin\frac{\theta_{0}}{2}\sinh\sigma
\label{udPmtre1}
\end{eqnarray}
We now consider the hyperbolic elements of SU(1,1) satisfying equation (~\ref{uhyp}).
Since now $| \alpha_{1}|>1$, setting
$\alpha_{1}=\epsilon\cosh\frac{\sigma}{2}$, $\epsilon =$ sgn $\lambda$,
  we obtain the eigenvalues as 
$\epsilon e^{\underline{+}\epsilon\frac{\sigma}{2}}$. Since the diagonal 
matrix
\begin{equation}\label{bst}
     \epsilon(\sigma) =\left(\begin{array}{cc}
     sgn \lambda e^{sgn \lambda (\frac{\sigma }{2})} &0\\
     0 &sgn \lambda e^{-sgn \lambda (\frac{\sigma }{2})}\end{array}\right)
\end{equation}
 belongs to SL(2,R), it can be regarded as the diagonal form of the
 matrix $g$ given by equations(~\ref{udHslD}) and (~\ref{udUSL}) with
$|\alpha_{1}|=\frac{|a+d|}{2}>1$. Henceforth we shall take $sgn \lambda = 1$.
The other case $sgn \lambda= -1 $ can be developed in an identical manner.

     An analysis parallel to the one for the elliptic elements shows
that for $(a+d)>2$ every matrix $g\in$SL(2,R) can be diagonalized also by
a matrix $v\in$SL(2,R). Thus 
\[
     g = v \epsilon(\sigma ) v^{-1}
\]
Since every matrix $v \in$ SL(2,R) can be decomposed as  
\[
     v=e(\theta)a(\rho)\epsilon(\alpha),\]
where 
\[     e(\theta)=\left(\begin{array}{cc}
     \cos\frac{\theta}{2} &\sin\frac{\theta}{2}\\
      -\sin\frac{\theta}{2} &\cos\frac{\theta}{2}
     \end{array}\right)\]
and $a(\rho)$ and $\epsilon(\alpha)$ are given by eqn (\ref{ast}) and (\ref{bst}
)
respectively. We therefore obtain the following parametrization of the
hyperbolic elements of $g\in$ SL(2,R),
\[
     g=e(\theta)a(\rho)\epsilon(\sigma)a^{-1}(\rho)e^{-1}(\theta)\]
The use of the isomorphism kernel in equation (~\ref{udISOslm}) then
yields 
\begin{eqnarray}\label{udPmtrh}
 \alpha= \cosh\frac{\sigma }{2}+i\sinh\frac{\sigma }{2}\sinh\rho
     \\
     \beta=-ie^{ - i\theta}\sinh\frac{\sigma }{2}\cosh\rho
\label{udPmtrh1}
\end{eqnarray}

     In Bargmann's theory $\!\!^{7}$ the carrier space for the discrete
series of representations of SU(1,1) was taken to be the functions
$\phi(z_{1},z_{2})$ where $z_{1},z_{2}$ are the spinors transforming
according to the fundamental representation of SU(1,1):
\[  (z_{1}^{\prime},z_{2}^{\prime})=(z_{1},z_{2})u\]  
 Since the fundamental representation of SU(1,1) and its complex conjugate
 are equivalent the functions $\phi(z_{1},z_{2})$ are required to satisfy 
\[\frac{\partial}{\partial \bar{z}_{1}}\phi(z_{1},z_{2})=
 \frac{\partial}{\partial \bar{z}_{2}}\phi(z_{1},z_{2})= 0,\]
 so that $\phi(z_{1},z_{2})$ is an analytic function of $z_{1}$ and $z_{2}$.
 The generators of SU(1,1) in this realization are given by,
\begin{eqnarray*}
      J_{1} &= &\frac{i}{2}\left(z_{1}\frac{\partial}{\partial
     z_{2}}+z_{2}\frac{\partial}{\partial z_{1}}\right)
     \\ 
      J_{2} &= &\frac{1}{2}\left(z_{1}\frac{\partial}{\partial
     z_{2}}-z_{2}\frac{\partial}{\partial z_{1}}\right)
     \\ 
      J_{3} &= &\frac{1}{2}\left(z_{1}\frac{\partial}{\partial
     z_{1}}-z_{2}\frac{\partial}{\partial z_{2}}\right) 
\end{eqnarray*}
where $J_{3}$ is the space rotation, and, $J_{1}$ and $J_{2}$ are pure
Lorentz boosts.\\
Explicit calculation yields,
\[J_{1}^{2}+J_{2}^{2}-J_{3}^{2}=K(1-K)\]
where,
\[ K=-\frac{1}{2}\left(z_{1}\frac{\partial}{\partial
     z_{1}}+z_{2}\frac{\partial}{\partial z_{2}}\right)\] 
Since $K$ commutes with $J_{1},J_{2}$ and $J_{3}$ it follows that in an
irreducible representation $\phi(z_{1},z_{2})$ is a homogeneous function
of degree $-2k$,
\[ \phi(z_{1},z_{2})=z_{2}^{-2k}f(z),\hspace{.4 in}
z=\frac{z_{1}}{z_{2}},\]
where $f(z)$ is an analytic function of $z$. It should be pointed out that
under the action of SU(1,1) the complex $z$-plane is foliated into three
orbits  a) $|z|<1$, b) $|z|>1$ and c) $|z|=1$ and the positive 
discrete series $D_{k}^{+}$ ( $ k = \frac{1}{2}, 1, \frac{3}{2}...$ ) is 
described by the first orbit $|z| < 1$, the open unit disc.  
Thus in Bargmann's construction the subspace $B_{k}(C)$ for 
$D_{k}^{+}$ consists of functions $f(z)$ analytic within the unit disc.

The finite element of the group in this realization can be easily obtained and is given by
\begin{equation}\label{urGRre}
      T_{u}f(z)=(\beta z + \bar{\alpha})^{-2k} f \left(
     \frac{\alpha z + \bar{\beta}}{\beta z + \bar{\alpha}} \right)
\end{equation}
These representations are unitary under the scalar product 
\begin{equation}\label{urpDEfn}
     (f,g)=\int_{|z|<1} \overline{f(z)}g(z) d\lambda(z)
\end{equation}
where 
\begin{eqnarray}\label{urpLAM}
     & &d\lambda(z)=\frac{(2k-1)}{\pi}(1-|z|^{2})^{2k-2}d^{2}z,\\
     & &z=x+iy, d^{2}z = dx dy \nonumber
\end{eqnarray}
The principal vector in $B_{k}(C)$ is given by
\begin{equation}\label{urvPvec}
     e_{z}(z_{1})=(1-\bar{z}z_{1})^{-2k}
\end{equation}
so that 
\begin{equation}\label{urprv} f(z)= \int_{|z_{1}|<1} (1-z\bar{z}_{1})^{-2k}f(z_{1})d\lambda(z_{1})\end{equation}
The action of the group ring, 
\[ T_{x}=\int x(u) T_{u} d\mu(u)\]
where $d\mu(u)$ is the invariant measure on SU(1,1) is given by
\begin{equation}\label{urfcc} 
 T_{x}f(z)=\int x(u)(\beta z+\bar{\alpha})^{-2k}f \left( \frac{\alpha
     z+\bar{\beta}}{\beta z+\bar{\alpha}} \right)d\mu(u)\end{equation}
Now, as before, using the basic property of the principal vector $i.e.$ (~\ref{urprv}) we have
\begin{equation}\label{urfnc}
 f \left( \frac{\alpha  z+\bar{\beta}}{\beta z+\bar{\alpha}} \right)=
     \int_{|z_{1}|<1} \left[ 1- \frac{(\alpha z+\bar{\beta})\bar{z}_{1}}{(\beta
     z+\bar{\alpha})} \right]^{-2k} f(z_{1})d\lambda(z_{1})\end{equation}
Substituting Eq. (~\ref{urfnc})
in Eq. (~\ref{urfcc}) we immediately obtain,
\[ T_{x}f(z)=\int_{|z_{1}|<1} K(z,z_{1})f(z_{1})d\lambda(z_{1})\]
where
\begin{equation}\label{ucK}
     K(z,z_{1})= \int x(u) (\beta z+ \bar{\alpha})^{-2k} \left[ 1- \frac{(\alpha z+\bar{\beta})\bar{z}_{1}}{(\beta
     z+\bar{\alpha})} \right]^{-2k} d\mu(u)
\end{equation}
Since the kernel, once again, is of the Hilbert-Schmidt type we have 
\[
 Tr(T_{x})= \int_{|z|<1}  K(z,z)d\lambda(z)
\]
Using Eq. (\ref{ucK}), the definition of the integral kernel, the above equation can be written in the form
\[  Tr(T_{x})=\int x(u) \pi(u) d\mu(u)\]
where the character $\pi(u)$ is given by 
\begin{equation}\label{2.80}
     \pi(u) = \int_{|z|<1} (\beta z+ \bar{\alpha})^{-2k} \left[ 1-
 \frac{(\alpha z+\bar{\beta})\bar{z}}{(\beta
     z+\bar{\alpha})} \right]^{-2k} d\lambda(z)
\end{equation}
We first consider the above integral for the elliptic elements of SU(1,1). 
Setting  
\begin{equation}\label{2.81}
 z= \tanh\frac{\tau}{2}e^{i\theta},\hspace{.2 in}
 0 \leq \tau < \infty, \hspace{.2 in} 0 \leq \theta \leq 2\pi
\end{equation}
and using the parametrization (\ref{udPmtre},\ref{udPmtre1}) for the elliptic elements we obtain after some calculations
\begin{equation}\label{2.82}
     \pi(u) =\frac{2k-1}{4\pi} \int_{\tau=0}^{\infty} d\tau\; \sinh\tau
 \int_{0}^{2\pi}d\phi \left[ \cos\frac{\theta_{0}}{2}
 -i\sin\frac{\theta_{0}}{2} (\cosh\sigma\cosh\tau +
 \sinh\sigma\sinh\tau\cos\phi) \right]^{-2k}
\end{equation}
where $\phi = \eta + \theta $. 
 We now introduce the time-like SO(2,1) unit vectors,
 \[\hat{n}=(-\sinh\sigma,0,\cosh\sigma), \hat{r}=(\sinh\tau\cos\phi
 ,\sinh\tau\sin\phi,\cosh\tau)\].
Then the Eq. (\ref{2.82}) can be written as 
\begin{equation}\label{ucC6}
     \pi(u) =\frac{2k-1}{4\pi} \int_{\tau=0}^{\infty} d\tau\; \sinh\tau
 \int_{0}^{2\pi}d\phi \left[ \cos\frac{\theta_{0}}{2}
 -i\sin\frac{\theta_{0}}{2} \hat{n}. \hat{r} \right]^{-2k}
\end{equation}
where $\hat{n}.\hat{r}$ stands for the Lorentz invariant form
 \[\hat{n}.\hat{r}=\hat{n}_{3}\hat{r}_{3}-\hat{n}_{2}\hat{r}_{2}
 -\hat{n}_{1}\hat{r}_{1}\]
Let us now perform a Lorentz transformation such that the time axis coincides with the fixed time-like SO(2,1) vector $\hat{n}$. Thus 
 \[\hat{n}.\hat{r}=\cosh\tau,\]
and we have
\[
\pi(u)=\frac{2k-1}{2} \int_{\tau=0}^{\infty}
  \left[ \cos\frac{\theta_{0}}{2}
 -i\sin\frac{\theta_{0}}{2} \cosh\tau \right]^{-2k} \sinh\tau d\tau\;\]
The above integration is quite elementary and it leads to Gel'fand and coworkers' formula for the character of the elliptic elements of SU(1,1)
 \[
 \pi(u)=\frac{e^{\frac{i}{2}\theta_{0}(2k-1)}}{e^{-i\frac{\theta_{0}}{2}}
 -e^{i\frac{\theta_{0}}{2}}}\]
 
For the hyperbolic elements of SU(1,1) we substitute the parametrization (\ref{udPmtrh},\ref{udPmtrh1}) in Eq. (\ref{2.80}) and use the transformation (\ref{2.81}). Thus
\begin{equation}\label{ucC6h}
     \pi(u) =\frac{2k-1}{4\pi} \int_{\tau=0}^{\infty} d\tau\; \sinh\tau
 \int_{0}^{2\pi}d\theta \left[ \cosh\frac{\sigma}{2}
 -i \hat{n}. \hat{r} \sinh\frac{\sigma}{2} \right]^{-2k}
\end{equation}
 where \[\hat{n}=(-\cosh\rho\cos\eta,-\cosh\rho\sin\eta,\sinh\rho)\] is a fixed 
space-like unit SO(2,1) vector and 
 \[\hat{r}=(\sinh\tau\cos\theta,\sinh\tau\sin\theta,\cosh\tau)\]
is a  time-like unit SO(2,1) vector. If we now perform a Lorentz transformation such that the first space axis (X-axis) coincides with the fixed space-like SO(2,1) vector $\hat{n}$ then
 \[\hat{n}.\hat{r}=\sinh\tau\cos\theta\]
so that
\begin{equation}\label{2.89}
 \pi(u) =\frac{2k-1}{4\pi} \int_{\tau=0}^{\infty} d\tau\; \sinh\tau
 \int_{\theta=0}^{2\pi}d\theta
  \left[ \cosh\frac{\sigma}{2}
 -i\sinh\tau \cos\theta \sinh\frac{\sigma}{2} \right]^{-2k}
\end{equation}
The evaluation of this integral is a little lengthy and is relegated to the appendix. Its value is given by
 \[
 \pi(u)=\frac{e^{-\frac{1}{2}\sigma(2k-1)}}{e^{\frac{\sigma}{2}}
 -e^{-\frac{\sigma}{2}}}\]

\section{The Principal Series of Representations}

  For the representations of the principal series we shall
realize the representations in the Hilbert space of functions $\!\!^{6}$ 
defined on the unit circle. For the representations of the
integral class
\[
T_u~ f(e^{i \theta}) = {\mid \beta e^{i \theta} + \bar{\alpha}
      \mid}^{-2k} ~ f\left( \frac{\alpha e^{i \theta} + \bar{\beta}}
               {\beta e^{i \theta} + \bar{\alpha}}\right) 
\]
For the representations of the half-integral class,
\[
T_u~ f(e^{i \theta}) = {\mid \beta e^{i \theta} + \bar{\alpha}
          \mid}^{-2k-1} ~ (\beta e^{i \theta} + \bar{\alpha}) 
            ~ f\left( \frac{\alpha e^{i \theta} + \bar{\beta}}
               {\beta e^{i \theta} + \bar{\alpha}}\right) 
\]
In both the cases
\[
k = \frac{1}{2} - is ~~~~~,~~~~~- \infty < s < \infty  
\]
In what follows we shall consider the integral class first.
For later convenience we replace $e^{i \theta}$ by
$exp[i(\theta - \pi /2)] = -ie^{i \theta}$.Thus
\[
T_u~ f(-ie^{i \theta}) = {\mid -i\beta e^{i \theta} + \bar{\alpha}
       \mid }^{-2k} ~ f\left( \frac{-i\alpha e^{i \theta} + \bar{\beta}}
               {-i\beta e^{i \theta} + \bar{\alpha}}\right) 
\]
We now construct the group ring
\[
T_x = \int x(u)~ T_{u}~ d \mu (u)
\]
so that
\[
T_x ~f(-ie^{i \theta}) = \int ~x(u)~ 
          {\mid -i\beta e^{i \theta} + \bar{\alpha}
      \mid }^{-2k}  ~f\left( \frac{-i\alpha e^{i \theta} + \bar{\beta}}
               {-i\beta e^{i \theta} + \bar{\alpha}}\right) 
                  ~d \mu (u)
\]
We now make a left translation
\[
u \rightarrow \underline{\theta} ^{-1} ~u
\]
where
\[
\underline{\theta} = \left(\begin{array}{cc}
                         e^{i \theta /2} & 0 \\
                         0 & e^{-i \theta /2}   
                     \end{array}\right)
\]
so that
\[
\alpha \rightarrow \alpha ~e^{-i \theta /2}~~,~~~~~
  \beta \rightarrow \beta ~e^{-i \theta /2}
\]
We therefore obtain
\[
T_x ~f(-ie^{i \theta}) = \int x(\theta ^{-1} u) 
         ~ {\mid -i\beta  + \bar{\alpha}
      \mid }^{-2k}  ~f\left( \frac{-i\alpha  + \bar{\beta}}
               {-i\beta  + \bar{\alpha}}\right) 
                  ~d \mu (u)
\]
We now map the $SU(1,1)$ matrix $u$ onto the $SL(2,R)$
matrix $g$ by using the isomorphism kernel $\eta$ given 
by eqns.(\ref{udISOslm}) and (\ref{udUSL}) and perform the Iwasawa              
decomposition 
\begin{equation}\label{3.9a}
g = k ~\theta _1
\end{equation}
where
\begin{equation}\label{3.9b}
k = \left(\begin{array}{cc}
         k_{11} & k_{12} \\
         0 & k_{22}
     \end{array}\right)
    ,~~~~~~~k_{11} k_{22} = 1
\end{equation}
belongs to the subgroup $K$ of real triangular marices
of determinant unity and $\theta _1 \in \Theta$ where $\Theta$  is the
subgroup of pure rotation matrices:
\begin{equation}\label{3.9c}
\theta _1 = \left(\begin{array}{cc}
             \cos (\theta _1 /2) & - \sin (\theta _1 /2) \\
             \sin (\theta _1 /2) & \cos (\theta _1 /2)
            \end{array}\right)
\end{equation}
We now introduce the following convention. The letters 
without a bar below it will indicate the $SL(2,R)$
matrices or its subgroups and those with a bar below it
will indicate their $SU(1,1)$ image. For instance
\[
\underline{k} = \eta ~k~ \eta ^{-1} = \frac{1}{2}
        \left(\begin{array}{cc}
         k_{11} + k_{22} - ik_{12}~~~~~ &
                     k_{12} - i(k_{11} - k_{22}) \\
         k_{12} + i (k_{11} - k_{22}) ~~~~~&
                     k_{11} + k_{22} + i k_{12}
        \end{array}\right)
\]
\[
\underline{\theta} _1 = 
        \left(\begin{array}{cc}
         e^{i \theta _1 /2} & 0 \\
         0 & e^ {-i \theta _1 /2}
        \end{array}\right)
\]
The decomposition (\ref{3.9a}) can also be written as
\begin{equation}\label{3.11}
u = \underline{k}~ \underline{\theta} _1
\end{equation}
which yields
\[
-i ~\alpha ~+~ \bar{\beta} ~=~ -i ~k_{22} ~e^{i \theta _1 /2}
\]
\[
-i ~\beta ~+~ \bar{\alpha} ~=~ k_{22} ~e^{-i \theta _1 /2}
\]
Hence setting $f(-ie^{i \theta}) = g(\theta)$ 
we obtain,
\begin{equation}\label{3.13}
T_x ~g(\theta) = \int 
     x(\underline{\theta} ^{-1} ~\underline{k}
            ~\underline{\theta} _1 )
    ~ {\mid k_{22} \mid }^{-2k} ~g (\theta _1)
     ~d \mu (u)
\end{equation}
It can be shown that under the decomposition
(\ref{3.9a},\ref{3.9b},\ref{3.9c}) or equivalently (\ref{3.11}) the invariant
measure decomposes as 
\begin{equation}\label{3.14}
d \mu (u) = \frac{1}{2} ~d \mu _l (g) = 
  \frac{1}{2} ~d \mu _r (g) =
     \frac{1}{4} ~d \mu _l (k) ~d \theta _1
\end{equation}
Substituting the decomposition (\ref{3.14}) in eq.
(\ref{3.13}) we have
\[
T_x ~g(\theta) = \int _{\Theta}
     K(\theta ,\theta _1) ~g(\theta _1)
       ~d \theta _1 
\]
where
\begin{equation}\label{3.15b}
K(\theta ,\theta _1) = \frac{1}{4} ~\int _{K}
   x(\underline{\theta} ^{-1} ~\underline{k}
       ~\underline{\theta} _1 )
   ~{\mid k_{22} \mid}^{-2k} 
      ~d \mu _l (k)
\end{equation}
Since the kernel $K(\theta , \theta _1)$ is of
the Hilbert-Schmidt type it has the trace
\[
Tr~(T_x) = \int _{\Theta } K(\theta ,\theta)
              ~d \theta
\]
Using the definition of the kernel as given by 
eq.(\ref{3.15b}) we have
\begin{equation}\label{3.16}
Tr~(T_x) = \frac{1}{4} ~\int _{\Theta ,K}
   x(\underline{\theta} ^{-1} ~\underline{k}
         ~\underline{\theta}) 
   ~{\mid k_{22} \mid }^{-2k} ~d \theta
     ~d \mu _l (k)
\end{equation}
Before proceeding any farther we note that 
$\underline{\theta}^{-1} ~\underline{k} ~\underline{\theta}$
represents a hyperbolic element of
$SU(1,1)$:
\[
u =
\underline{\theta} ^{-1} ~\underline{k} ~\underline{\theta}
\]
Calculating the trace of both sides we have
\begin{equation}\label{3.17b}
k_{22} + 1 / k_{22} = 2 \alpha _1
\end{equation}
In the previous section we have seen that for
the elliptic elements of $SU(1,1)$ \\ $\alpha _1
= \cos(\theta _0 /2) < 1 $. The eqn.(\ref{3.17b}),
therefore, for the elliptic case yields
\[
k_{22} ^2 - 2 k_{22} \cos (\theta _0 /2) + 1 = 0
\]
which has no real solution. Thus the elliptic
elements of $SU(1,1)$ do not contribute to the
character of the principal series of representations.
We, therefore, assert that for this particular 
class of unirreps the trace is concentrated on
the hyperbolic elements.

    We shall now show that every hyperbolic
element of $SU(1,1)$ ( i.e those with \\ 
$\mid\alpha_1\mid=\mid(a+d)\mid/2>1$
) can be
represented as
\begin{equation}\label{3.18a}
u = 
\underline{\theta} ^{-1} ~\underline{k} ~\underline{\theta}
\end{equation}
or equivalently as 
\begin{equation}\label{3.18b}
g = \theta ^{-1} ~k ~\theta
\end{equation}
Here $k_{11} = \lambda ^{-1}$, $k_{22} = \lambda$
are the eigenvalues of the matrix $g$ taken in 
any order.

    We recall that every $g \in SL(2,R)$ for the 
hyperbolic case can be diagonalized as
\[
v' ~g ~v'^{-1} = \delta
\]
where
\[
\delta = \left(\begin{array}{cc}
          \delta _1 & 0 \\
          0 & \delta _2
         \end{array}\right)
  ,~~~~~\delta _1 \delta _2 = 1;~~~~~
   \delta _1 , \delta _2 ~~~~real 
\]
belongs to the subgroup $D$ of real diagonal
matrices of determinant unity and \\
$v'\in SL(2,R)$.If we write the Iwasawa decomposition
for $v'$
\[
v' = k' ~\theta
\]
then 
\[
g = \theta ^{-1} ~{k'}^{-1} ~\delta ~k' ~\theta
\]
Now ${k'}^{-1} ~\delta ~k' \in K$ so that writing
$k = {k'} ^{-1} ~\delta ~k'$ we have the decomposition
(\ref{3.18b}) in which
\[
k_{11} = \delta _1 = \lambda ^{-1} ,~~~~
    k_{22} = \delta _2 = \lambda
\]

    If these eigenvalues are distinct then for
a given ordering of them the matrices $k$,
$\theta$ are determined uniquely by the 
matrix $g$. In fact we have
\[
k_{12} = b-c , ~~~~~ \tan \theta =
 \frac{(a-d)(b-c) + (\lambda - \lambda ^{-1})(b+c)}
     {(b-c)(b+c) + (\lambda - \lambda ^{-1})(a-d)}
\]
It follows that for a given choice of $\lambda$
the parameters $\theta$ and $k_{12}$ are
uniquely determined. We note that there are 
exactly two representations of the matrix $g$
by means of formula (\ref{3.18b}) corresponding
to two distinct possibilities
\[
k_{11} = sgn \lambda {\mid \lambda \mid}^{-1} =
  sgn \lambda ~e^{\sigma /2},~~~~~
k_{22} = sgn \lambda ~\mid \lambda \mid
     = sgn \lambda ~e^{- \sigma /2}
\]
\[
k_{11} = sgn \lambda \mid \lambda \mid =
  sgn \lambda ~e^{- \sigma /2},~~~~~
k_{22} = sgn \lambda ~\mid \lambda \mid
     = sgn \lambda ~e^{\sigma /2}
\]
Let us now remove from $K$ the elements with
$k_{11} = k_{22} =1$. This operation cuts 
the group $K$ into two connected disjoint
components. Neither of these components 
contains two matrices which differ only by
permutation of the two diagonal elements.
In correspondence with this partition the 
integral in the r.h.s. of eqn.(\ref{3.16}) is
represented in the form of a sum of two
integrals
\begin{equation}\label{3.24}
Tr~(T_x) =\frac{1}{4} \int _ {\Theta} d\theta \int _ {K_1}
  d \mu _l (k) ~{\mid k_{22} \mid}^{-2k}
  ~x(
\underline{\theta} ^{-1} ~\underline{k} ~\underline{\theta})
 ~+~ \frac{1}{4} \int _{\Theta} d \theta \int _{K_2}
d \mu _l (k) ~{\mid k_{22} \mid }^{-2k} ~x(
\underline{\theta} ^{-1} ~\underline{k} ~\underline{\theta})
\end{equation}
As $\theta$ runs over the subgroup $\Theta$
and $k$ runs over the components $K_1$ or
$K_2$ the matrix \\ $g = \theta ^{-1} ~k~ \theta$
runs over the hyperbolic elements of the
group $SL(2,R)$ or equivalently \\ $u =  
\underline{\theta} ^{-1} ~\underline{k} ~\underline{\theta}$
runs over the hyperbolic elements of the
$SU(1,1)$. We shall now prove that in $K_1$
or $K_2$
\begin{equation}\label{3.25}
d \mu _l (k) ~d \theta = \frac{4 \mid k_{22} \mid }
   {\mid k_{22} - k_{11} \mid} ~d \mu (u)
\end{equation}
To prove this we start from the left invariant
differential element
\[
dw = g^{-1} ~dg
\]
where $g \in SL(2,R)$ and $dg$ is the matrix of the 
differentials $g_{pq}$ i.e. following the 
notation of eqn.(\ref{udHslD})
\[
dg = \left(\begin{array}{cc}
       da & db \\
       dc & dd 
     \end{array}\right)
\]
The elements $dw$ are invariant under the left
translation $g \rightarrow g_0 g$. Hence
choosing a basis in the set of all $dg$ we 
immediately obtain a differential left invariant
measure. For instance choosing $dw_{12}$, 
$dw_{21}$, $dw_{22}$ as the independent invariant
differentials we arrive at the left invariant   
measure on $SL(2,R)$.
\begin{equation}\label{3.27}
d \mu _l (g) = dw_{12} dw_{21}  dw_{22}
\end{equation}
In a similar fashion we can define the right
invariant differentials
\[
dw' = dg~g^{-1}
\]
which is invariant under the right translation
$g \rightarrow g g_0$.

   To prove the formula (\ref{3.25}) we write the
decomposition (\ref{3.18b}) as
\[
\theta ~g = k ~\theta
\]
so that
\begin{equation}\label{3.30}
d \theta ~g + \theta ~dg = dk ~\theta + k ~d \theta
\end{equation}
From eqn.(\ref{3.30}) it easily follows that
\begin{equation}\label{3.31}
dw = g^{-1} ~dg = \theta ^{-1} ~d \mu ~\theta
\end{equation}
where
\begin{equation}\label{3.32}
d \mu = k^{-1} ~dk + d \theta ~\theta ^{-1} ~-
   ~k^{-1} ~d \theta ~\theta ^{-1} ~k
\end{equation}
In accordance with the choice of the independent
elements of $dw$ as mentioned above we choose
the independent elements of $d \mu$ as $d \mu _{12}$,
$d \mu _{21}$, $d \mu _{22}$. The eqn.(\ref{3.31}) then
leads to
\begin{equation}\label{3.33a}
dw_{11} + dw_{22} = d \mu _{11} + d \mu _{22}
\end{equation}
\begin{equation}\label{3.33b}
dw_{11} dw_{22} - dw_{12} dw_{21} = d\mu_{11} d\mu_{22}
      - d\mu_{12} d\mu_{21}
\end{equation}
Further since $Tr(d \mu) = Tr(dw) =0$ we immediately
obtain from eqn.(\ref{3.33a},\ref{3.33b})
\[
dw_{22}^2 + dw_{12} dw_{21} = d\mu_{22}^2 + d\mu_{12}
     d\mu_{21}
\]
which can be written in the form
\begin{equation}\label{3.35}
d\eta_1^2 + d\eta_2^2 - d\eta_3^2 = 
   {d\eta'}_1^2 + {d\eta'}_2^2 - {d\eta'}_3^2 
\end{equation}
where
\[
\begin{tabular}{ll}
 $d \eta_1 = (dw_{12} + dw_{21}) /2,~~~~~~~~~~~~~~$  & $d \eta '_1 = 
            (d \mu_{12} + d \mu_{21}) /2$  \\
 $d \eta_2 = dw_{22}$,  & $ d \eta ' _2 = d \mu _{22}$  \\
 $d \eta_3 = (dw_{12} - dw_{21}) /2$, & $ d \eta ' _3 =
            (d \mu_{12} - d \mu_{21}) /2 $ 
\end{tabular}
\]
The eqn.(\ref{3.35}) implies that the set $d \eta$ and the 
set $d \eta '$ are connected by a Lorentz transformation.
Since the volume element $d \eta _1 d \eta _2 d \eta _3 $
is invariant under such a transformation we have,
\begin{equation}\label{3.37}
d \eta _1 d \eta _2 d \eta _3 = 
   d \eta ' _1 d \eta ' _2 d \eta ' _3
\end{equation}
But the l.h.s. of eqn.(\ref{3.37}) is $dw_{12} dw_{21} dw_{22} /2$ 
and r.h.s. is $d\mu_{12} d\mu_{21} d\mu_{22} /2$. 
Hence using eqn.(\ref{3.27}) we easily obtain
\[
d\mu_l (g) = d\mu_{12} d\mu_{21} d\mu_{22}
\]
We now write eqn.(\ref{3.32}) in the form
\begin{equation}\label{3.39}
d\mu = du + dv
\end{equation}
where $du = k^{-1} dk$ is the left invariant differential
element on $K$ and
\begin{equation}\label{3.40}
dv = d \theta ~\theta ^{-1} ~-~ k^{-1} ~d \theta ~\theta ^{-1} ~k
\end{equation}
In eqn.(\ref{3.39}) $du$ is a triangular matrix whose
independent nonvanishing elements are chosen to be
$du_{12}$, $du_{22}$ so that
\[
d\mu _l (k) = du_{12} du_{22}
\]
On the other hand $dv$ is a $2 \times 2$ matrix having
one independent element which is chosen to be $dv_{21}$.
Since the Jacobian connecting $d \mu _{12} d \mu _{21} d
\mu _{22}$ and $du_{12} du_{22} dv_{21}$ is a triangular
determinant having $1$ along the main diagonal we
obtain
\begin{equation}\label{3.42}
d\mu _l (g) = d\mu_l (k)~dv_{21}
\end{equation}
It can now be easily verified that each element $k
\in K$ with distinct diagonal elements (which is
indeed the case for $K_1$ or $K_2$) can be represented          
uniquely in the form
\begin{equation}\label{3.43}
k = \zeta ^{-1} ~\delta ~\zeta
\end{equation}
where $\delta$ belongs to the subgroup of real
diagonal matrices with unit determinant and $\zeta \in Z$,
where Z is a subgroup of $K$ consisting of real matrices
of the form
\[
\zeta = \left(\begin{array}{cc}
          1 & \zeta _{12} \\
          0 & 1 
        \end{array}\right)
\]
Writing eqn.(\ref{3.43}) in the form $\zeta ~k = \delta ~\zeta$ we
obtain
\begin{equation}\label{3.45}
k_{pp} = \delta _p ~~,~~~~~~~~\zeta _{12} = k_{12} /
       (\delta _1 - \delta _2 )
\end{equation}
Using the decomposition (\ref{3.43}) we can now write eqn.(\ref{3.40})
in the form
\[
dv = \zeta ^{-1} ~dp ~\zeta
\]
where
\[
dp = d \lambda ~-~ \delta ^{-1} ~d \lambda ~\delta
\]
\[
d \lambda = \zeta ~d \sigma ~\zeta ^{-1}~~~~~, ~~~~~
   d \sigma = d \theta ~\theta ^{-1}
\]
From the above equations it now easily follows
\begin{equation}\label{3.47}
dv_{21} = \frac{\mid \delta _2 - \delta _1 \mid}
   {\mid \delta _2 \mid} ~\frac{d \theta}{2}
\end{equation}
Substituting eqn.(\ref{3.47}) in eqn.(\ref{3.42}) and using eqns.(\ref{3.14})
and (\ref{3.45}) we immediately obtain eqn.(\ref{3.25}).

       Now recalling that in $K_1$, $\mid k_{22} \mid = e^{- \sigma /2}$
and in $K_2$, $\mid k_{22} \mid = e^{\sigma /2}$ the eqn.(\ref{3.24})
in conjunction with (\ref{3.25}) yields
\[
Tr~(T_x) = \int x(u) ~\pi (u) ~d \mu (u)
\]
where the character $\pi (u)$ is given by
\[
\pi (u) = \frac{e^{(2k-1) \sigma /2} + e^{-(2k-1) \sigma /2}}
     {\mid e^{\sigma /2} - e^{-\sigma /2} \mid}
\]
For the principal series of representations belonging to
the half-integral class a parallel calculation yields
\[
\pi (u) = \frac{e^{(2k-1) \sigma /2} + e^{-(2k-1) \sigma /2}}
     {\mid e^{\sigma /2} - e^{-\sigma /2} \mid} ~sgn \lambda 
\]

\appendix
\section{Evaluation of the integral ( \ref{2.89}) }

By a simple change of variable the eq.(\ref{2.89}) can be written in the
form, 
\begin{eqnarray*}
\pi(u) = \left[ \frac{(2k - 1)}{2 \pi} \right] \int_{0}^{\infty} \sinh\tau d \tau
\int_{0}^{\pi} d \theta \left[ \cosh \left(\frac{\sigma}{2}\right) - i
\sinh \left(\frac{\sigma}{2}\right) \cos \theta \sinh \tau \right]^{-2k}
\end{eqnarray*}
To carry out the $\theta$ integration we set
\[ x = \frac{1}{2} ( 1 - \cos \theta), \hspace{.3 in} 0 \leq x \leq 1
\]
and use the formula $\!\!^{8}$ 
\[
F(a,b;c;z) = \frac{\Gamma (c)}{\Gamma(b) \Gamma(c-b)} \int_{0}^{1}
x^{b-1} (1-x)^{c-b-1} (1-zx)^{-a} dx, \hspace{.3 in} Re(c) > Re(b)> 0
\]
so that
\begin{eqnarray*}
\pi(u) = \left[ \frac{(2k - 1)}{2 } \right] \int_{0}^{\infty} \sinh\tau d \tau
\left[ \cosh \left(\frac{\sigma}{2}\right) - i
\sinh \left(\frac{\sigma}{2}\right) \sinh \tau \right]^{-2k} \nonumber\\  
F\left(2k,\frac{1}{2},1;\frac{-2i \sinh\frac{\sigma}{2}\sinh\tau}{\left[\cosh\left(\frac{\sigma}{2}\right)-i\sinh\frac{\sigma}{2}\sinh\tau\right]}\right)
\end{eqnarray*}
To evaluate the above integral we use the quadratic transformation $\!\!^{9}$ 
\[
F(a,b;2b;z) = \left(1-\frac{z}{2}\right)^{-a}F\left(\frac{a}{2},\frac{1}{2}+\frac{a}{2};b+\frac{1}{2};\left[\frac{z}{2-z}\right]^2 \right)
\]
Thus 
\[
\pi(u) = \left[\frac{(2k-1)}{2} \right] \int \left[\cosh \left(\frac{\sigma}{2} \right)
\right]^{-2k} F\left(k,\frac{1}{2}+k;1;-\tanh^{2}\left(\frac{\sigma}{2}\right)\sinh^{2}\tau\right)
\sinh\tau d\tau
\]
For integral $k$ we now extract the branch point of the hypergeometric function by 
using $\!\!^{10}$
\[
F(a,b;c;z) = (1-z)^{-b}F(c-a,b;c;\frac{z}{z-1})
\]
Using the above formula we obtain after some calculations
\begin{eqnarray*}
\pi(u) = \left[\frac{(2k-1)}{2} \right] \left[\cosh \left(\frac{\sigma}{2} \right)
\right]^{-2k}
\left[(sech^{2}~
\left(\frac{\sigma}{2} \right)
\right]^{-\frac{1}{2}-k} ~~~\sum_{n=0}^{k-1}  
\frac{(1-k)_{n} (\frac{1}{2}+k)_{n}}{(1)_{n}n!}  \left[\sinh^{2}\left(
\frac{\sigma}{2} \right)\right]^{n}\nonumber\\
\int_{0}^{\infty}  \left[1 + \sinh^{2} \left(\frac{\sigma}{2} \right)
\cosh^{2}\tau \right]^{-\frac{1}{2}-k-n}
(\cosh^{2}\tau-1)^{n}\sinh\tau d \tau \;\;
\end{eqnarray*}
Setting $\cosh\tau = t^{-\frac{1}{2}}$ we obtain after some calculations
\begin{eqnarray}
\pi(u) = \left[\frac{(2k-1)}{4} \right] \left[\cosh \left(\frac{\sigma}{2} \right)\right]
\left[\sinh^2 \left(\frac{\sigma}{2} \right)
\right]^{-\frac{1}{2}-k} \Gamma(k) \sum_{n=0}^{k-1} 
\frac{(1-k)_{n} (\frac{1}{2}+k)_{n}}{n!\Gamma(n+k+1)}\nonumber \\
F\left(\frac{1}{2}+k+n, k; n+k+1; -cosech^{2} \left(\frac{\sigma}{2} \right)\right)
\label{A8}  \end{eqnarray}
The summation over $n$ can be carried out by expanding the hypergeometric function 
appearing in Eq. (\ref{A8}) in a power series. Thus,
\begin{eqnarray*}
\pi(u) = \left[\frac{(2k-1)}{4} \right] \left[\cosh \left(\frac{\sigma}{2} \right)\right]
\left[\sinh^2 \left(\frac{\sigma}{2} \right)
\right]^{-\frac{1}{2}-k} \frac{\Gamma(k)}{\Gamma(\frac{1}{2}+k)}  \sum_{n=0}^{\infty}
\frac{(k)_{n} (\frac{1}{2}+k+n)}{n!\Gamma(n+k+1)} \nonumber \\
 \left[ -cosech^{2} \left(\frac{\sigma}{2} \right) \right]^{n} 
F\left(1-k, \frac{1}{2}+k+n; n+k+1; 1\right)
\end{eqnarray*}
The hypergeometric function of unit argument can be summed by Gauss's formula $\!\!^{11}$ and we obtain after some calculations
\[ \pi(u)= \cosh \left(\frac{\sigma}{2} \right)\left[(\sinh^2 \left(\frac{\sigma}{2} \right)
\right]^{-\frac{1}{2}-k} (2)^{-2k}
F\left(k,k+\frac{1}{2};2k;- cosech^{2} \left(\frac{\sigma}{2} \right)\right)\]
We now use the formula $\!\!^{12}$ 
\[ F\left(k,k+\frac{1}{2}; 2k;z \right) = (1-z)^{-\frac{1}{2}} \left[\frac{1}{2}(1+\sqrt{1-z})\right]
^{(1-2k)} \]
which immediately yields
 \[
 \pi(u)=\frac{e^{-\frac{1}{2}\sigma(2k-1)}}{e^{\frac{\sigma}{2}}
 -e^{-\frac{\sigma}{2}}}\]

for half-integral $k$ a parallel calculation yields the same result.


\begin{references}

 $^1$ I.M.Gel'fand and M.A. Naimark; {\em I.M.Gel'fand - Collected Papers, Vol 
I$\!$I}, p 41, 182  ( Springer-Verlag, Berlin-Heidelberg 1988) \\
 
 $^2$ See ref. 1, p.41.\\

 $^3$     I.M.Gel'fand, M.I.Graev and I.I.Pyatetskii--Shapiro; {\em Representation
Theory and Automorphic Functions, Chapter-2}, p 199,202 (W.B. Saunders,
Philadelphia, London, Toronto; 1969)\\

 $^4$     V. Bargmann, Commun. Pure. Appl. Math. {\bf 14}, 187 (1961); 
{\bf 20}, 1 (1967); in {\em Analytic Methods in Mathematical Physics, edited by
P. Gilbert and R. G. Newton} (Gordon and Breach, New York, 1970) \\

 $^5$     I.E. Segal, Ill.J.Math {\bf 6}, 500 (1962)\\

 $^6$  N. Ja. Vilenkin and A. U. Klymik, {\em Representations of Lie groups and 
special functions, Vol. 1, Chap. 6}, p. 298 (Kluwer Academic, Boston.1991)\\

 $^7$     V. Bargmann, Ann.Math. {\bf 48}, 568 (1947)\\

 $^8$ A. Erdelyi (ed.), {\em Higher Transcendental Functions, Bateman Manuscript 
Project, Vol. 1}\\ 
( McGraw Hill, New York, 1953),  p. 114.\\

 $^9$  See ref. 8, p.111.\\

 $^{10}$  See ref. 8, p.109.\\

 $^{11}$  See ref. 8, p.104.\\


 $^{12}$  See ref. 8, p.101.\\

\end{references}
\end{document}